\documentstyle[12pt]{article}
\newcommand{\be}{\begin{equation}}
\newcommand{\ee}{\end{equation}}
\newcommand{\bea}{\begin{eqnarray}}
\newcommand{\eea}{\end{eqnarray}}
\newcommand{\norsl}{\normalsize\sl}
\newcommand{\norsc}{\normalsize\sc}
\begin{document}

\begin{titlepage}

\title{The Mass of the Lightest Supersymmetric Higgs Boson
       beyond the Leading Logarithm Approximation}

\author{
\norsc  Jiro KODAIRA\thanks{Work partially supported by
          the Monbusho Grant-in-Aid for Scientific Research
          (C) No. 05640351.} \,and Yoshiaki YASUI
         \thanks{Fellow of the Japan Society for the Promotion of
         Science for Japanese Junior Scientist. Work partially supported
         by the Monbusho
          Grant-in-Aid for Scientific Research No. 050076.}\\
\norsl  Dept. of Physics, Hiroshima University\\
\norsl  Higashi-Hiroshima 724, JAPAN\\
\\
\norsc  Ken SASAKI\thanks{e-mail address: a121004@c1.ed.ynu.ac.jp}\\
\norsl  Dept. of Physics, Yokohama National University\\
\norsl  Yokohama 240, JAPAN\\}

\date{}
\maketitle

\begin{abstract}
{\normalsize We examine the radiative corrections to the mass of
the lightest Higgs boson in the minimal supersymmetric extension of the
standard model.
We use the renormalization-group improved effective potential which includes
the next-to-leading-order contributions.  We find that, contrary to the result
of Espinosa and Quir\'os, the higher-order corrections to the lightest
Higgs boson mass are non-negligible, adding $3-11$ GeV ($3-9$ GeV) to the
result in the leading logarithm approximation for the range of top
quark mass $100 {\rm GeV} < m_t < 200 {\rm GeV}$ and for
the supersymmetric breaking scale $M_{\rm SUSY} = 1 {\rm TeV}$
($M_{\rm SUSY} = 10 {\rm TeV}$). Also we find that our result is stable
under the change of the renormalization parameter $t$.}
\end{abstract}

\begin{picture}(5,2)(-290,-610)
\put(2.3,-110){HUPD-9316}
\put(2.3,-125){YNU-HEPTh-93-102}
\put(2.3,-140){November 1993}
\end{picture}

\thispagestyle{empty}
\end{titlepage}
\setcounter{page}{1}
\baselineskip 18pt

Although the standard model (SM) is highly successful and
in excellent agreement with all the measurements at the present energies,
it is widely believed that SM is not the final theory
for the world of elementary particles. The minimal supersymmetric
extention of the standard model (MSSM) is one of the most promising candidates
beyond the SM. The MSSM possesses in its physical spectrum three neutral
and two charged Higgs bosons, and there exists a tree-level
relation  which implies  that at least one neutral
Higgs boson is lighter than the $Z^0$ mass ($M_Z$).
Radiative corrections
to the masses of
these Higgs bosons have been calculated by several
groups~\cite{rO} -~\cite{rLS},
who found that they are quite significant,
depending strongly on the top quark
mass and the scale of supersymmetry breaking
($M_{\rm SUSY}$) or the squark masses.
All the above works except ref.~\cite{rEQ} considered the one-loop radiative
corrections to Higgs boson masses.

Now that the one-loop corrections have been found to be significantly
large, it is quite natural to ask next how large the higher-order
corrections would be. Indeed Espinosa and Quir\'os~\cite{rEQ} have
analyzed the \lq\lq two-loop\rq\rq \,
radiative corrections to the mass of the lightest Higgs boson in the
minimal and non-minimal (including a gauge singlet) supersymmetric
standard model. They used the effective potential (EP) in the leading
logarithm approximation and examined the evolution of the Higgs
quartic coupling $\lambda$ by the
renormalization group (RG) techniques with the one- and two-loop
$\beta$ functions.
They found that the \lq\lq two-loop\rq\rq \,
correction is negative and stays within a few percent even in cases
where the one-loop correction is larger than the tree-level
mass.

Recently there appeared interesting papers~\cite{rKa} -~\cite{rEin}
which discussed about the
improvement of EP by using the
renormalization group equation (RGE) .
It was shown there that to improve EP
which satisfies the RGE
with up to the two-loop $\beta$ functions and anomalous dimension $\gamma$,
one should include the one-loop-level potential with the
running parameters into the solution. In this respect, the work of
Espinosa and Quir\'os~\cite{rEQ} seems unsatisfactory: they used the
EP in the leading logarithm approximation, which is the
tree-level potential with the running parameters, and they made use
of the two-loop
$\beta$ functions only to determine the evolution of these parameters.
In this paper we reanalyze the mass
of the lightest Higgs boson $(m_{\phi})$ in the MSSM using the EP
improved by RGE up to the next-to-leading order. We find that new terms
which were not considered by
Espinosa and Quir\'os give non-negligible contributions to the $m_{\phi}$.
We also find that the predicted values of $m_{\phi}$ are stable under the
change of the renormalization parameter $t$ when we use the RGE-improved
EP which includes the next-to-leading-order contributions.

Two basic assumptions were made in their analysis
of the lightest Higgs boson mass~\cite{rEQ}:
(a) all supersymmetric (SUSY)
partners of the SM particles have masses of the order of the supersymmetry
breaking scale $M_{\rm SUSY}$; (b) one linear combination $H$ of
the two Higgs boson doublets, $H_1=(H_1^0, H_1^{-})^T$ and
$H_2=(H_2^{+}, H_2^0)^T$,
\be
  H = H_1 {\rm cos}\beta + i\tau _{2} H_2^{\ast} {\rm sin}\beta
\ee
is light, while
the other linear combination, which is orthogonal to the former one,
is as heavy as the SUSY partners.  Under these assumptions,
it is clear that the effective theory below the scale $M_{\rm SUSY}$ is
the usual SM with one light Higgs doublet $H$. Throughout the following
analyses, we will make the same
assumptions (a) and (b).
The tree-level Higgs potential below $M_{\rm SUSY}$ is then given by
\be
  V_{\rm tree} = -m^2 | H |^2 + \frac{1}{6} \lambda | H |^4 ,
\label{Vtree}
\ee
where
\be
    \frac{1}{3} \lambda = \frac{1}{4}(g_1^2+g_2^2){\rm cos}^2 2\beta,
\label{Lamb}
\ee
and $g_1$ and $g_2$ are the gauge coupling constants of
$U(1)_Y$ and $SU(2)_L$, respectively.

When the neutral component of $H$
acquires a vacuum expectation value ${<H^0>}= v/\sqrt{2}$,
the above tree-level potential (\ref{Vtree}) gives the physical
Higgs $\phi \equiv ({\rm Re}H^0 - v)/\sqrt{2}$
(which corresponds to the lightest Higgs boson
under the assumptions (a) and (b)) a squared mass
\be
   m^2_{\phi} = \frac{1}{3} \lambda v^2.
\label{Hmass}
\ee
Also at the symmetry breaking, the top quark and the $Z^0$, $W^{\pm}$
gauge bosons
gain masses which are given by
\be
      m_t = h_t v/\sqrt{2}, \quad M^2_Z = \frac{1}{4}(g_1^2+g_2^2) v^2,
 \quad  M^2_W = \frac{1}{4} g_2^2 v^2,
\label{tmass}
\ee
where $h_t$ is the Yukawa coupling
of $\phi$ to the top quark. The tree-level relation
$m_{\phi}^2=M_Z^2 {\rm cos}^2 2\beta $ \,follows from
Eqs.~(\ref{Lamb}) -~(\ref{tmass}).

The (RGE-unimproved) EP of the SM up to the one-loop level
is given by
\bea
      V_1 &=&V_{(0)} + V_{(1)}, \label{EV} \\
     V_{(0)} &=& -\frac{1}{2} m^2 \varphi _{c}^2
                +\frac{1}{24} \lambda \varphi _{c}^4, \label{EVa} \\
     V_{(1)} &=& -\frac{3}{64\pi ^2}(h_t^2 \varphi _{c}^2)^2
       ({\rm ln} \frac{h_t^2 \varphi _{c}^2}{2\mu ^2} - \frac{3}{2})
        + \cdots \cdots,
\label{EVb}
\eea
where $\varphi _{c}$ is the classical field
corresponding to the physical Higgs boson $\phi$, and all the Yukawa couplings
of $\phi$ to quarks and leptons except the top quark are
neglected. The calculation is performed in the Landau gauge and
in the $\overline{MS}$ scheme to obtain the one-loop result $V_{(1)}$, and
$\mu$ is the renormalization scale. The ellipses in Eq.~(\ref{EVb})
represent contributions of the gauge bosons and the would-be-Goldstone
bosons. Throughout this paper we use the Landau gauge
which is the most convenient for our purpose~\cite{rCW}
and the $\overline{MS}$ scheme.

Now we improve the EP by using the RGE. It was recently emphasized
by the authors of
Refs.~\cite{rKa} -~\cite{rEin} that in the $\overline{MS}$ scheme
the EP $V(\varphi _{c})$ fails to satisfy the usual
(homogeneous) RGE unless $V(0)$, a contribution
to the \lq\lq vacuum energy\rq\rq, is suitably
dealt with. When we use the RGE-improved EP
in the leading order and obtain $m_{\phi}$, the consideration of
$V(0)$ term is unnecessary.
However, as we shall see below,
if we improve the EP by RGE up to the next-to-leading
order, $V(0)$ becomes relevant to us and we must take its presence
into account. Thus with an appropriate $\varphi _{c}$-independent
term being added, the new EP $V(\varphi _{c})$ satisfies the following RGE
of the usual form:
\be
        ({\cal D} -
         \gamma _{\phi} \varphi _{c} \frac{\partial}{\partial \varphi _{c}})
            V(\varphi _{c}, X_i, \mu ) = 0
\label{RGE}
\ee
with
\be
      {\cal D}=\mu \frac{\partial}{\partial \mu} +
         \beta _{X_i} \frac{\partial}{\partial {X_i}} ,
\label{Der}
\ee
where $X_i = \lambda, h_t, m^2, g_3, g_2, g_1$ and $g_3$ is the gauge coupling
constant of $SU(3)_C$. The solution is easily found by the method of
characteristics and we get
\be
        V(\varphi _{c}, X_i, \mu )
         = V(\varphi _{c} (t), X_i (t), \mu (t) ),
\label{Sola}
\ee
where $X_i (t)$ are running couplings and running mass
which are determined by the equations
\be
       \frac{dX_i (t)}{dt} = \beta _{X_i} (X_j (t)),
            \qquad X_i,X_j = \lambda, h_t, m^2, g_3, g_2, g_1
\label{tdepa}
\ee
with the boundary conditions $X_i (0)=X_i$ , and
\bea
       \xi (t) &=& {\rm exp} \left\{ -\int_{0}^{t} \gamma _{\phi} (t^\prime)
                                 dt ^\prime \right\},\nonumber \\
    \varphi _{c} (t) &=& \varphi _{c} \xi (t), \label{tdepb} \\
            \mu (t) &=& \mu e^t.\nonumber
\eea
Then using the result of tree- and one-loop-level EP
of Eqs.(\ref{EVa}) and (\ref{EVb}), we obtain the RGE-improved $V$ as follows:
\bea
             V &=& \Omega (X_i (t), \mu (t) ) \nonumber\\
                & &+ V_{(0)}(\varphi _{c} (t),
                          X_i (t))\nonumber\\
                & &+ V_{(1)} (\varphi _{c} (t),
                    X_i (t), \mu (t) ) + \cdots  ,\label{Solb}
\eea
where $\Omega$ is the $\varphi _{c}$-independent
term which is added for $V$ to satisfy
a RGE of the usual homogeneous form,
and the ellipses represent the higher-loop contributions.

For later convenience, let us expand the RGE coefficient functions
$\beta_{X_i}$ and $\gamma_{\phi}$ by the number of loops as follows:
\bea
        \beta _{X_i} &=& \hbar \beta _{X_i}^{(1)}
                     +\hbar ^2 \beta _{X_i}^{(2)} + \cdots,
                \qquad X_i=\lambda, h_t, m^2, g_3, g_2, g_1 \nonumber\\
      \gamma_{\phi} &=& \hbar \gamma_{\phi}^{(1)}
             + \hbar ^2 \gamma _{\phi}^{(2)} + \cdots,
\eea
where we have introduced the Plank's constant $\hbar$ so that
the power of $\hbar$ counts the number of loops and
$\beta _{X_i}^{(n)}$ and $\gamma_{\phi}^{(n)}$ are
the n-loop contribution
to $\beta _{X_i}$ and $\gamma_{\phi}$, respectively.
Similarly $V$ has the loop expantion
\be
           V= \Omega + V_{(0)}(\varphi _{c}(t))
           + \hbar V_{(1)}(\varphi _{c}(t)) + \cdots ,
\label{Exb}
\ee
and we have denoted $\Omega (X_i (t), \mu (t) )$ and
$V_{(n)} (\varphi _{c} (t), X_i (t), \mu (t) )$
as $\Omega$ and $V_{(n)}(\varphi _{c}(t))$, for short, respectively.

Inserting Eq.(\ref{Exb}) into Eq.(\ref{RGE}) and picking the terms up to
of the order $\hbar$, we find
\be
     {\cal D}\Omega
      +\hbar \left\{ \beta _{X_{i}}^{(1)} \frac{\partial V_{(0)}}{\partial X_i}
               -\gamma _{\phi}^{(1)} \varphi_c
                 \frac{\partial V_{(0)}}{\partial \varphi_c}
                +{\frac{3}{32\pi ^2}} h_t ^4 \varphi_c ^4 \right\} =0,
\label{Omega}
\ee
where the last term in the parentheses arises from
$\mu {\partial V_{(1)}}/{\partial \mu}$ when we use the expression of $V_{(1)}$
in Eq.(\ref{EVb}) and neglect the contributions of the gauge bosons and
the would-be-Goldstone bosons to $V_{(1)}$.
If we further set $\varphi_c =0$ in the above equation, we obtain
\be
        {\cal D}\Omega
           +\hbar \mu \frac{\partial V_{(1)}(\varphi_c =0)}{\partial \mu} =0,
\label{Omega2}
\ee
from which $\Omega$ can be determined to the leading order.
For the later discussion, however, we do not need the specific form of
$\Omega$ , and we will see below that the knowledge
of Eq.(\ref{Omega}) is sufficient for our purpose.

We will now analyze the mass of the lightest Higgs boson using the RGE-improved
$V$. At first, let us consider the boundary conditions for
coupling constants. Under our basic assumptions explained before, the
relation between the quartic coupling constant $\lambda$ and the gauge
coupling constants $g_1$ and $g_2$ given in Eq.(\ref{Lamb}) should be
satisfied
at the scale $M_{\rm SUSY}$, that is,
\be
      \frac{1}{3} \lambda (M_{\rm SUSY}) =
                \frac{1}{4}(g_1^2(M_{\rm SUSY}) + g_2^2(M_{\rm SUSY}))
                 {\rm cos}^2 2\beta.
\label{Lam}
\ee
So we choose the renormalization scale
$\mu$ to be $M_{\rm SUSY}$ and take the parameter $t$ as
$t={\rm ln}\left( \varphi _{c}/M_{\rm SUSY}\right)$.
Then we find that $\mu (t)=\varphi _{c}$ and the RGE-improved $V$ is given by
\be
           V= \Omega + V_{(0)}(\varphi _{c}(t))
           + \hbar V_{(1)}(\varphi _{c}(t)) + {\cal O}(\hbar ^{2}) ,
\label{VVVa}
\ee
with
\be
      V_{(1)}(\varphi _{c}(t))=
          -\frac{3}{64\pi ^2}h_t^4 (t) \varphi _{c}^4 (t)
       \left[{\rm ln} \frac{h_t^2 (t) \xi ^2 (t)}{2} - \frac{3}{2}\right]  ,
\label{VVVb}
\ee
where in $V_{(1)}(\varphi _{c}(t))$ we only include the top-quark contribution
to the one-loop EP,
because, with its very large Yukawa coupling, the contribution of top-quark is
dominant over those from the gauge bosons and the would-be Goldstone bosons.
We will obtain the lightest Higgs boson mass
by evaluating ${\partial ^{2} V}/{\partial {\varphi _{c}(t)^2}}$ at
$\varphi _{c}(t_{v}) = v = (\sqrt{2}G_{\rm F})^{-\frac{1}{2}}
= 246 {\rm GeV}$ under the minimum condition
${\partial V}/{\partial {\varphi _{c}(t)}}=0$ at $\varphi _{c}(t_{v}) = v$.
The value $t_v$ is determined
by the equation $\varphi _{c}(t_{v}) = v$, which, with help of
Eq.(\ref{tdepb}),
is transformed into
\be
            t_v ={\rm ln} \frac{v}{M_{\rm SUSY}} +
                   \int_{0}^{t_v} \gamma _{\phi} (t^\prime)
                                 dt ^\prime.
\label{tv}
\ee
It is noted that we differentiate $V$ not by
$\varphi _{c}$ but by the renormalized field
$\varphi _{c}(t)$ and also that we evaluate the differetials at the point of
$\varphi _{c}(t_{v}) = v$ and not at $\varphi _{c}= v$.

Since $X_{i}(t)$, $\varphi _{c}(t)$, and $\mu (t)$ are functions of $t$,
we find
\bea
        \frac{\partial X_{i}(t)}{\partial {\varphi _{c}(t)}} &=&
           \hbar \beta _{X_{i}}^{(1)}(t)\frac{1}{\varphi _{c}(t)} +
                      {\cal O}(\hbar ^{2})  \nonumber\\
        \frac{\partial \mu (t)}{\partial {\varphi _{c}(t)}} &=&
              \frac{\mu (t)}{\varphi _{c}(t)} + {\cal O}(\hbar ).
\label{deriva}
\eea
Thus we obtain
\bea
     \frac{\partial \Omega}{\partial {\varphi _{c}(t)}} &=&
        \frac{1}{\varphi _{c}(t)}
        \left\{\mu (t) \frac{\partial \Omega}{\partial {\mu (t)}} +
         \hbar \beta _{X_{i}}^{(1)}(t)
            \frac{\partial \Omega}{\partial {X_{i}(t)}}\right\} +
               {\cal O}(\hbar ^{2}),  \nonumber \\
     \frac{\partial  V_{(0)}(\varphi _{c}(t))}{\partial {\varphi _{c}(t)}} &=&
         V'_{(0)}(\varphi _{c}(t)) +
               \hbar \frac{\beta _{X_{i}}^{(1)}(t)}{\varphi _{c}(t)}
            \frac{\partial  V_{(0)}(\varphi _{c}(t))}{\partial {X_{i}(t)}} +
           {\cal O}(\hbar ^{2}),  \label{derivb} \\
     \frac{\hbar \partial  V_{(1)}(\varphi _{c}(t))}{\partial {\varphi
_{c}(t)}}
        &=&-\hbar \frac{3}{16\pi ^2} h_t^4 (t) \varphi _{c}^3 (t)
       \left[{\rm ln} \frac{h_t^2 (t) \xi ^2 (t)}{2} - \frac{3}{2}\right]
        + {\cal O}(\hbar ^{2}). \nonumber
\eea
It is now straightforward to evaluate $\partial V/{\partial {\varphi _{c}(t)}}$
from Eq.(\ref{derivb}). Using the relation (\ref{Omega}), we eliminate
${\cal D}\Omega$ term and find
\bea
    \frac{\partial V}{\partial {\varphi _{c}(t)}}
       &=&(1+\hbar \gamma _{\phi}^{(1)}(t))
            \left\{-m^{2}(t)\varphi _{c}(t)+
              \frac{1}{6}\lambda (t)\varphi _{c}^{3}(t)\right\} \nonumber \\
 & & +\hbar \left\{-\frac{3}{16\pi ^{2}} h_{t}^{4}(t)\varphi _{c}^{3}(t)
               \left[{\rm ln}\frac{h_{t}^{2}(t)\xi ^{2}(t)}
                           {2} -1 \right]\right\} \nonumber \\
       & & +{\cal O}(\hbar ^{2}).
\label{Vd}
\eea
Further differentiation of $\partial V/{\partial {\varphi _{c}(t)}}$ by
$\varphi _{c}(t)$ gives
\bea
    \frac{\partial ^{2}V}{\partial {\varphi _{c}^{2}(t)}}
       &=&(1+\hbar \gamma _{\phi}^{(1)}(t))
          \left\{-m^{2}(t) +
           \frac{1}{2}\lambda (t)\varphi _{c}^{2}(t)\right\} \nonumber \\
       & &+ \hbar\left\{-\beta _{m^{2}}^{(1)}(t)
           +\frac{1}{6}\beta_{\lambda}^{(1)}(t)\varphi _{c}^{2}(t)\right\}
                    \nonumber \\
       & &+\hbar\left\{-\frac{9}{16\pi ^{2}} h_{t}^{4}(t)
                    \varphi _{c}^{2}(t)
                \left[{\rm ln}\frac{h_{t}^{2}(t)\xi^{2}(t)}
                           {2} -1 \right]\right\} \nonumber \\
       & & +{\cal O}(\hbar ^{2}).
\label{Vdd}
\eea
Using the minimum
condition $\partial V/\partial {\varphi _{c}(t)}=0$ at
$\varphi _{c}(t_{v})=v$, we finally obtain for the lightest Higgs mass
in the next-to-leading logarithm approximation
\bea
 m_{\phi ({\rm 2-loop})}^2
    &=& \left. \frac{\partial ^{2}V}{\partial {\varphi _{c}^{2}(t)}}
           \right | _{\varphi _{c}(t_{v}) = v}\nonumber \\
    &=& \frac{1}{3} \lambda (t_{v}) v^2 \nonumber \\
    & &+ \hbar v^2 \Bigl\{ \frac{1}{6}
                 \beta_{\lambda}^{(1)}(t_v)
               - \frac{1}{6} \lambda (t_v) \Bigl [
    \frac{\beta _{m^2}^{(1)}(t_v)}{m^{2}(t_v)}-2\gamma _{\phi}^{(1)}(t_v)
                            \Bigr ] \label{massNL} \\
    & &\qquad \qquad - \frac{3}{8\pi ^2}h_t(t_v)^4 \Bigl [
           \ln \frac{h_t^2(t_v)\xi ^2(t_v)}{2} - 1
                            \Bigr ]\Bigr \} \nonumber \\
    & &+ {\cal O}(\hbar ^2). \nonumber
\eea

The first term in Eq.(\ref{massNL}) gives
\be
       m_{{\phi}{({\rm 1-loop})}}^2 =(1/3)\, \lambda (t_v) v^2,
\label{massL}
\ee
which is the result given by
Refs.~\cite{rO} \cite{rBFC} except that the authors of Ref.~\cite{rO}
have evaluated the running coupling $\lambda (t)$ at $t_{m_{\phi}}
= \ln (m_{\phi}/M_{\rm SUSY})$.
As far as the leading logarithm approximation is concerned,
the terms of order $\hbar$ in Eq.(\ref{massNL}) are neglected as
the higher-order effects. An arbitrariness coming from different choice of
the parameter $t$ also falls in the higher-order corrections although
it has no small effect on the predicted values numerically.
Espinosa and Quir\'os have employed the \lq\lq one-loop\rq\rq \,formula
$m_{\phi}^2 = (1/3)\, \lambda (t) v^2$, which is correct only
in the leading logarithm approximation, and computed $\lambda (t)$
at $t_{m_{\phi}} = \ln (m_{\phi}/M_{\rm SUSY})$ using
RGE with up to two-loop $\beta$ functions~\cite{rEQ}. However, if we make use
of the two-loop RGE coefficient functions for
the running parameters and evaluate
the lightest Higgs mass $m_{\phi}^2$, we should take into
account the order-$\hbar$-terms in Eq.(\ref{massNL})
which also collect the next-to-leading
logarithmic contributions.

The one- and two-loop $\beta$ functions
and anomalous dimension $\gamma_{\phi}$ for the SM,
which we will use in this analysis, read as follows~\cite{rEin}.
We define the constant $A$ as $A\equiv 16\pi ^2$.

\noindent For the Higgs quartic coupling $\lambda$:
\bea
    A\beta _{\lambda}^{(1)} &=& 4\lambda ^2
  + 12\lambda h_t^2 -36h_t^4 -3\lambda (3g_2^2 +g_1^2)
  + \frac{27}{4}g_2^4 +\frac{9}{2}g_2^2 g_1^2 +\frac{9}{4}g_1^4 ,
    \nonumber\\
  A^2\beta _{\lambda}^{(2)} &=& -\frac{26}{3}\lambda ^3 -24\lambda ^2
         h_t^2 +6\lambda ^2 (3g_2^2 +g_1^2)\nonumber\\
   & & +\lambda \{-3h_t^4+h_t^2 (80g_3^2 +\frac{45}{2}g_2^2+
       \frac{85}{6}g_1^2)-\frac{73}{8}g_2^4 +\frac{39}{4}g_2^2 g_1^2
       +\frac{629}{24}g_1^4\}\nonumber\\
   & & +180h_t^6 -h_t^4 (192g_3^2 +16g_1^2 )+h_t^2(-\frac{27}{2} g_2^4
  +63g_2^2 g_1^2 -\frac{57}{2}g_1^4 )\label{quarbeta} \\
   & & +\frac{915}{8}g_2^6 -\frac{289}{8}g_2^4 g_1^2-\frac{559}{8}g_2^2
       g_1^4-\frac{379}{8}g_1^6.\nonumber
\eea

\noindent For the top-quark Yukawa coupling $h_t$:
\bea
   A\beta _{h_t}^{(1)} &=& \frac{9}{2}h_t^3-
      h_t(8g_3^2+\frac{9}{4}g_2 +\frac{17}{12}g_1^2),\nonumber\\
 A^2\beta _{h_t}^{(2)} &=& h_t\Bigl\{-12h_t^4 -2\lambda h_t^2+
          h_t^2 (36g_3^2+\frac{225}{16}g_2^2+\frac{131}{16}g_1^2)\label{topY}\\
  & & +\frac{1}{6}\lambda^2 -108g_3^4 +9g_3^2 g_2^2 +\frac{19}{9}g_3^2
   g_1^2-\frac{23}{4}g_2^4 -\frac{3}{4}g_2^2 g_1^2 +
      \frac{1187}{216}g_1^4 \Bigr \}.\nonumber
\eea

\noindent For the gauge couplings $g_3$, $g_2$, and $g_1$:
\bea
   A\beta _{g_3}^{(1)} &=& -7g_3^3,\nonumber\\
 A^2\beta _{g_3}^{(2)} &=& g_3^3 (-2h_t^2 -26g_3^2
       +\frac{9}{2}g_2^2 +\frac{11}{6}g_1^2),\nonumber\\
   A\beta _{g_2}^{(1)} &=& -\frac{19}{6}g_2^3,\label{gaugebeta}\\
 A^2\beta _{g_2}^{(2)} &=& g_2^3 (-\frac{3}{2}h_t^2 +12g_3^2
          +\frac{35}{6}g_2^2 +\frac{3}{2}g_1^2),\nonumber\\
   A\beta _{g_1}^{(1)} &=& \frac{41}{6}g_1^3,\nonumber\\
 A^2\beta _{g_1}^{(2)} &=& g_1^3 (-\frac{17}{6}h_t^2 +\frac{44}{3}g_3^2
        +\frac{9}{2}g_2^2 +\frac{199}{18}g_1^2).\nonumber
\eea

\noindent For the mass parameter $m^2$:
\bea
   A\beta _{m^2}^{(1)} &=& m^2 (2\lambda +6h_t^2
       -\frac{9}{2}g_2^2-\frac{3}{2}g_1^2 ),\nonumber\\
 A^2\beta _{m^2}^{(2)} &=& m^2 \{-\frac{5}{3}\lambda^2-12\lambda h_t^2
       +4\lambda (3g_2^2 +g_1^2)-\frac{27}{2}h_t^4\label{mbeta}\\
    & & +h_t^2 (40g_3^2 +\frac{45}{4}g_2^2 +\frac{85}{12}g_1^2)
      -\frac{145}{16}g_2^4+\frac{15}{8}g_2^2 g_1^2
         +\frac{157}{48}g_1^4\}.\nonumber
\eea

\noindent For the anomalous dimension $\gamma_{\phi}$:
\bea
    A\gamma _{\phi}^{(1)} &=& 3h_t^2-\frac{9}{4}g_2^2-\frac{3}{4}g_1^2,
                  \nonumber\\
 A^2 \gamma _{\phi}^{(2)} &=& \frac{1}{6}\lambda ^2-\frac{27}{4}h_t^4
       +h_t^2 (20g_3^2+\frac{45}{8}g_2^2+\frac{85}{24}g_1^2 )\label{anom}\\
    & & -\frac{271}{32}g_2^4 +\frac{9}{16}g_2^2 g_1^2
                 +\frac{431}{96}g_1^4 .\nonumber
\eea

When we substitute the expressions of the RGE coefficient functions
$\beta _{\lambda}^{(1)}$, $\beta _{m^2}^{(1)}$, and $\gamma _{\phi}^{(1)}$
into Eq.(\ref{massNL}), we find that many terms cancel out and we obtain
a rather simple expression
\bea
   m_{\phi ({\rm 2-loop})}^2
     &=& \frac{1}{3} \lambda v^2 \nonumber \\
     & &+ \hbar v^2 \frac{1}{16\pi ^2}\Bigl\{ \frac{1}{3}\lambda^{2}
             +2\lambda h^{2}_t -\frac{1}{2} \lambda (3g_2^2+g_1^2)
                 \label{massNLA} \\
     & &\qquad \qquad \qquad +\frac{9}{8}g_2^4 +\frac{3}{4}g_2^2 g_1^2
                 +\frac{3}{8}g_1^4 - 6h_t^4
           \ln \frac{h_t^2\xi ^2}{2}
                            \Bigr \} \nonumber
\eea
where it is understood that all the running parameters are the ones
evaluated at $t=t_v$. It is interesting to note, in particular, that
a $h^4_t$ term in Eq.(\ref{massNL}), which comes from the one-loop EP
of Eq.(\ref{EVb}), cancels
with another $h^4_t$ term in $\beta _{\lambda}^{(1)}$ of Eq.(\ref{quarbeta})
and, in consequence, there appears only one $h^4_t$ term of the form
$h^4_t{\rm ln}(h^2_t \xi^2/2)$ in the order-$\hbar$ contributions in
Eq.(\ref{massNLA}). This makes the higher-order
corrections to $m_{\phi ({\rm 2-loop})}^2$ to be a milder one. From the
expression of Eq.(\ref{massNLA}) we expect that the next-to-leading-order
corrections give a positive contribution to $m_{\phi}^2$, which will be shown
numerically to be true below.

To evaluate $m_{\phi ({\rm 2-loop})}$ of Eq.(\ref{massNLA}) numerically,
we choose the initial conditions for the gauge couplings
$\alpha _i \equiv g_i^2 /{4\pi}$
(i=3,2,1) at the scale $M_Z =91.2 {\rm GeV}$ to be
\be
  \alpha _3 (M_Z)=0.115, \quad \alpha _2 (M_Z)=0.0336, \quad
       \alpha _1 (M_Z)=0.0102
\label{data}
\ee
which are consistent with present experimental constraints~\cite{rEKN}
-~\cite{rAm}, and define the Yukawa coupling of the top quark at the
scale of its mass $m_t$ as
\be
   h_t (m_t)= \sqrt{2} m_t /v \qquad {\rm with} \quad v=246 {\rm GeV}.
\label{topm}
\ee
The fact that the two-loop $\beta_{\lambda}^{(2)}$,
$\beta_{h_t}^{(2)}$, and $\beta_{g_i}^{(2)}$ (i=3,2,1) are functions of
the couplings $\lambda$, $h_t$, and $g_i$
casts  Eq.(\ref{tdepa}) into a very complicated system of coupled
differential equations.
For given values of $m_t$ and $M_{\rm SUSY}$, we
first solve the system (\ref{tdepa}) and (\ref{tv})
with $\beta _{X_i}$ and $\gamma _{\phi}$ given in Eqs.(\ref{quarbeta})
-~(\ref{anom}) together with the
initial conditions Eqs.(\ref{data}) and (\ref{topm}), and we obtain the
appropriate $t_v$ and $\lambda (t_v)$  so that $\lambda$ satisfies the boundary
condition, Eq.(\ref{Lam}), when
it evolves from $t_v$ to $t=0$.
At the same time when we find the appropriate $t_v$ and $\lambda (t_v)$, we
gain
all the information on the parameters which appear in the r.h.s. of
Eq.(\ref{massNLA}).
This is how we calculate $m_{\phi ({\rm 2-loop})}$ for
given values of $m_t$ and $M_{\rm SUSY}$.

In Fig.1 we plot $m_{\phi ({\rm 2-loop})}$ as a function of $m_t$
for $M_{\rm SUSY} = 1\, {\rm TeV}$,
$\cos ^2 2\beta = 1$ and $\cos ^2 2\beta = 0$
along with $m_{\phi ({\rm 1-loop})}$ in the
leading logarithm approximation. Fig.2 shows the case for
$M_{\rm SUSY} = 10\, {\rm TeV}$. Since it is suggested that
$m_t$ is not too excessively large in MSSM ~\cite{rDSB}, we have studied
$m_{\phi ({\rm 2-loop})}$ for $m_t$
from $100 {\rm GeV}$ to $200 {\rm GeV}$. The curves for $\cos ^2 2\beta = 1$
($\cos ^2 2\beta = 0$) can be considered as upper (lower) bounds for the
lightest Higgs boson mass in the MSSM.
 From Fig.1 and 2 we observe that the next-to-leading-order effects
are non-negligible. They add $3-11$ GeV ($3-9$ GeV) to the
result in the leading logarithm approximation for the range of top
quark mass $100 {\rm GeV} < m_t < 200 {\rm GeV}$ and for
$M_{\rm SUSY} = 1 {\rm TeV}$  ($M_{\rm SUSY} = 10 {\rm TeV}$).
These rather large corrections come from the order-$\hbar$-terms of
Eq.(\ref{massNLA}), since without those terms we could indeed recover
the result of Ref.~\cite{rEQ}, namely, the higher-order corrections
being negative and negligible for the considered range of parameters.
Contrary to the conclusion of Espinosa and Quir\'os, our result shows that
the higher-order corrections turn out to be positive and
non-negligible when top quark is very heavy.

It is to be noted
that our result in the leading logarithm approximation differs
numerically from those of Refs.~\cite{rO}~\cite{rEQ} because we
evaluated the running coupling $\lambda (t)$ at $t_v$ given by Eq.(\ref{tv}),
instead of at $t_{m_{\phi}}= \ln (m_{\phi}/M_{\rm SUSY})$.
In other words, we have made a different choice
of $t$ from the ones made in the above references.
In the case of $\cos ^2 2\beta = 1$, for example, our predicted values for
$m_{\phi ({\rm 1-loop})}$ are smaller than those calculated in
Refs.~\cite{rO}~\cite{rEQ} by $0 - 6{\rm GeV}$ ($0 - 10{\rm GeV}$) for the
range $100 {\rm GeV} < m_t < 200 {\rm GeV}$ and for
$M_{\rm SUSY} = 1 {\rm TeV}$ ($M_{\rm SUSY} = 10 {\rm TeV}$).
In the leading logarithm approximation, the predicted values for
the lightest Higgs boson mass
are rather sensitive to the choice of the renormalization parameter $t$.
However, in the next-to-leading
logarithm approximation, it is not possible
to change the definition of $t$ without modifying the ${\cal O}(\hbar)$
terms of EP. In consequence, the result is stable under the change of $t$.
This is a well-known issue which often arises in the
renormalization group approach, and will be discussed in more detail
in the second comment below.

Fig.1 shows that the next-to-leading-order corrections are large for
$\cos ^2 2\beta = 0$ especially when $M_{\rm SUSY} = 1 {\rm TeV}$.
In the case of $\cos ^2 2\beta = 0$, the boundary condition
for $\lambda (t)$ at $M_{\rm SUSY}$ is
$\lambda (t=0)=0$. When $M_{\rm SUSY} = 1 {\rm TeV}$, the ``evolution time"
is not long enough for $\lambda (t)$ to grow from the initial value
$\lambda =0$, and thus a term of the form $-6h_t^4 \ln (h_t^2\xi ^2/2) $
and gauge-coupling-constant terms in the order-$\hbar$ contributions in
Eq.(\ref{massNLA}) give relatively large corrections compared with
the leading $(1/3)\lambda (t_v) v^2$.

A few comments are in order.
The first comment concerns the definition of mass.
The \lq\lq mass\rq\rq \,
we have calculated is not the on-shell mass. It might be necessary to
consider the correction coming from the wave function renormalization
in order to make a realistic prediction for the Higgs mass.
However, this effect is
expected to be small~\cite{rERT}~\cite{rLSZ}.

Secondly, we have taken the parameter $t$ as $t={\rm ln}
(\varphi _{c}/M_{\rm SUSY})$ to derive Eq.(\ref{massNL}). But
the physics should not depend on the choice of $t$.
For example, as was stressed in Refs.~\cite{rKu}~\cite{rEin}, the
\lq\lq natural choice\rq\rq \,of $t$ may be given by the equation
\be
  2\mu ^2 (t) = 2\mu ^2 e^{2t} = h_t ^2 (t) \varphi _c^2 (t).
\label{newt}
\ee
Then the RGE-improved EP which we deal with will be
\bea
  \widetilde{V} &=& \Omega (X_i (t), \mu (t) )
                    -\frac{1}{2}m^2 (t) \varphi _c^2 (t)
                    +\frac{1}{24}{\lambda (t)}\varphi _{c}^4(t) \nonumber \\
          & &+\hbar \left \{\frac{9}{128\pi ^2}h_t(t)^4 \varphi _{c}^4(t)
      \right \}.
\label{newV}
\eea
Since the expression of the order-$\hbar$ term of the above equation is
different from the one we analysed before,
we may think at first sight that we would obtain a different result
for $m_{\phi ({\rm 2-loop})}$. In fact we follow the same procedure as before,
i.e., first differentiate $\widetilde{V}$ by $\varphi _{c}(t)$, use
the relation (\ref{Omega}) and eliminate ${\cal D}\Omega$ term, evaluate
${\partial ^{2} \widetilde{V}}/{\partial {\varphi _{c}(t)^2}}$ at
$\varphi _{c}(\widetilde{t_{v}}) = v$ under the minimum condition
${\partial \widetilde{V}}/{\partial {\varphi _{c}(t)}}=0$ at
$\varphi _{c}(\widetilde{t_{v}}) = v$, and we obtain the following expression:
\bea
 \widetilde{m}_{\phi ({\rm 2-loop})}^2
    &=& \frac{1}{3} \lambda (\widetilde{t_v}) v^2 \nonumber \\
    & &+ \hbar v^2 \Bigl\{ \frac{1}{6}
                 \beta_{\lambda}^{(1)}(\widetilde{t_v})
               - \frac{1}{6} \lambda (\widetilde{t_v}) \Bigl [
    \frac{\beta _{m^2}^{(1)}(\widetilde{t_v})}
        {m^{2}(\widetilde{t_v})}-2\gamma _{\phi}^{(1)}(\widetilde{t_v})
                            \Bigr ] \label{massNLc} \\
    & &\qquad \qquad + \frac{3}{8\pi ^2}h_t(\widetilde{t_v})^4
        \Bigr \} \nonumber \\
      & &+ {\cal O}(\hbar ^2). \nonumber
\eea
A logarithmic term of the form $-(3/{8\pi ^2})\hbar v^2
h_t^4{\rm ln}(h^2_t \xi^2/2)$ which appeared in Eq.(\ref{massNL}) is
missing. However, we should note that the definition of $t$ has been altered.
Remembering that we have chosen the renormalization scale $\mu$ to
be $M_{\rm SUSY}$, we must now evaluate the
curvature of EP at the value of $\widetilde{t_v}$, which is
determined by
\be
    \widetilde{t_v} = {\rm ln} \frac{v}{M_{\rm SUSY}} +
                 \ln \frac{h_t(\widetilde{t_v})}{ \sqrt{2}}.
\label{scale}
\ee
Expanding $\lambda (\widetilde{t_v})$ around $t_v$, we find
\bea
    \lambda (\widetilde{t_v}) &\approx& \lambda (t_v) +
       \hbar \beta_{\lambda}^{(1)}(t_v) \left[\widetilde{t_v}-t_v \right]
                    \nonumber  \\
         &\approx& \lambda (t_v) -\hbar \frac{9}{8\pi ^{2}} h_t^{4}(t_v)
                  {\rm ln}\frac{h_{t}^{2}(t_v)\xi ^{2}(t_v)}{2}
          + \cdots \cdots.
\label{newlamb}
\eea
Then substituting the above expression for $\lambda (\widetilde{t_v})$ into
Eq.(\ref{massNLc}), we will obtain essentially the same result as
Eq.(\ref{massNL}) except for the $h_t^2$- and other non-dominant terms.
In fact, we have calculated the mass $\widetilde{m}_{\phi ({\rm 2-loop})}$
with the above choice of $t$, Eq.(\ref{newt}), and found
the difference between the two numerical results
for $m_{\phi ({\rm 2-loop})}$ and $\widetilde{m}_{\phi ({\rm 2-loop})}$
being less than $1 {\rm GeV}$ for $m_t < 200 {\rm GeV}$
and $M_{\rm SUSY} < 10 {\rm TeV}$.

In conclusion we have examined the mass of the lightest Higgs boson
in the MSSM beyond the leading logarithm approximation. We have made
use of the EP improved by RGE up to the next-to-leading order. We have
found that, contrary to the result of Espinosa and Quir\'os, the
next-to-leading-order corrections to the Higgs mass are non-negligible,
adding $3 - 11 {\rm GeV}$ ($3 - 9 {\rm GeV}$) to the values predicted
by the RGE approach in the leading logarithm approximation for the
range $100 {\rm GeV} < m_t < 200 {\rm GeV}$ and for
$M_{\rm SUSY} = 1 {\rm TeV}$ ($M_{\rm SUSY} = 10 {\rm TeV})$.
We also found that the predicted values of $m_{\phi}$ are stable under the
change of the renormalization parameter $t$ when we use the RGE-improved
EP which includes the next-to-leading-order contributions.

\vspace{2cm}
\begin{center}
{\large\bf Acknowledgements}
\end{center}
\medskip
We would like to thank M. Bando, T.Kugo, N. Maekawa, H. Nakano
and Y. Okada for discussions and useful comments. K.S. would like to
thank Bill Marciano and Rob Pisarski for discussions and the hospitality
extended to him at Brookhaven National Laboratory in the summer of 1993,
when part of this work was done.


\newpage
\noindent
{\large\bf Figure caption}
\medskip

\noindent
Fig.1

\noindent
Values of the Higgs boson mass as a function of $m_t$, for
$M_{\rm SUSY} = 1 {\rm TeV}$ and $\cos ^2 2\beta = 1$ (upper two lines)
and $\cos ^2 2\beta = 0$ (lower two lines).
The solid and dash-dotted lines denote the next-to-leading-order and the
leading-order results, respectively.

\medskip

\noindent
Fig.2

\noindent
The same as in Fig. 1, but considering the case for
$M_{\rm SUSY} = 10 {\rm TeV}$.

\end{document}